\documentclass[12pt]{article}
\usepackage{graphics}
\usepackage{amssymb}

\newcommand{\D}{\mathrm{d}}
\newcommand{\e}{\mathrm{e}}

\begin{document}

\title{Spectra of soft ring graphs}
\author{P.~Exner$^{a,b}$ and M.~Tater$^a$}
\date{}
\maketitle
\begin{quote}
{\small \em a) Department of Theoretical Physics, Nuclear Physics
Institute, \\ \phantom{e)x}Academy of Sciences, 25068 \v Re\v z,
Czech Republic \\
 b) Doppler Institute, Czech Technical University, B\v{r}ehov{\'a} 7,\\
\phantom{e)x}11519 Prague, Czech Republic \\
 \rm \phantom{e)x}exner@ujf.cas.cz, tater@ujf.cas.cz}
\vspace{8mm}

\noindent {\small We discuss of a ring-shaped soft quantum wire
modeled by $\delta$ interaction supported by the ring of a
generally nonconstant coupling strength. We derive condition which
determines the discrete spectrum of such systems, and analyze the
dependence of eigenvalues and eigenfunctions on the coupling and
ring geometry. In particular, we illustrate that a random
component in the coupling leads to a localization. The discrete
spectrum is investigated also in the situation when the ring is
placed into a homogeneous magnetic field or threaded by an
Aharonov-Bohm flux and the system exhibits persistent currents.}
\end{quote}

%%%%%%%%%%%%%%%%%%%%%%%%%%%%%%%%%%%%%%%%%%%%%%%%%%%%%%%%%%%%%%%%%%%

\section{Introduction}

The aim of the present paper is to investigate spectral properties
of a two-dimensional quantum particle subject to a $\delta$
interaction supported by a circular ring, plus possibly a magnetic
field. Similar systems were studied for a quite a long time. One
can mentioned an early paper by Kurilev \cite{Ku} and later work
\cite{BT, BEKS} devoted to measure-type perturbations of the
Laplacian where some general spectral results were
derived\footnote{Recall also a related model of a photonic crystal
studied in \cite{FK, KuK}. It differs from the present by the
switched roles of the coupling and spectral parameters.}. Later
this line of study was extended to the situation when a singular
zero-measure set itself supports a nontrivial dynamics \cite{Kar,
KK, Ko}.

In the most part of the mentioned work the emphasis was laid on
methods to treat such singular interactions. Recently the problem
was put into a different perspective related to a specific
physical model describing idealized ``quantum wires''. These
tube-like or graph-like semiconductor structures are usually
modeled as strips or tubes with hard walls, or in a more
simplified version by graphs supporting ``one-dimensional''
electrons. A weak point of such models is the assumption about the
nature of the confinement. In actual quantum wires, the electrons
are trapped due to interfaces between two different semiconductor
materials which represents a finite potential jump. Thus electrons
can be found also outside such a wire, although not ``too far''
since the exterior represents here a classically forbidden region.

Recent investigations revealed interesting relations between the
spectrum and geometry of such a ``soft'' quantum wire, namely the
existence of curvature-induced bound states \cite{EI}. Moreover,
it was demonstrated that in the strong-coupling limit the negative
part of the spectrum of such Hamiltonians approaches that of an
``ideal'' curve-like wire with an effective potential determined
by the curvature \cite{EY1, EY2}. On the other hand, apart of a
few very simple examples we are lacking solvable models of such
systems.

In the present paper we analyze a simple model with a $\delta$
interaction supported by a circle of radius $R$ and a generally
non-constant coupling strength. In particular, we are interested
in the discrete spectrum of such ring systems. First we illustrate
that it is rather the geometry of the interaction curve than its
topology which determines the spectral properties. To this aim we
discuss in Section~2 the case of the ``full'' and ``broken'' ring,
the latter being obtained by putting the coupling constant zero at
its segment; we will show that the spectrum depends substantially
on the ability of the particle to tunnel though the ``gap''.
Another natural question concerns spectral properties in the
situation when the coupling strength is randomly varying. It is
customary to investigate it in an infinite-length setting as a
passage from absolutely continuous to dense pure point spectrum
\cite{St}. At the same time the localization effect can be seen in
systems with perturbations of a finite length too if we observe
the characteristic size of the wavefunction; we will illustrate
this fact in the framework of the present model.

Another situation we are going to discuss concerns the case when
the particle is exposed to a magnetic field perpendicular to the
ring plane. In Section~3 we discuss two models, one with a
homogeneous background field and the other with a pointlike
Aharonov-Bohm flux piercing the centre of the ring. For an ideal
wire these situations are, of course, equivalent because the only
quantity which counts is the flux through the loop. This is no
longer true in the soft case when the particle wave packet extends
outside the ring.

An important question concerns the existence of current-carrying
states. Recall that persistent currents in rings threaded by a
magnetic flux are one of the characteristic features of mesoscopic
systems -- see, e.g., \cite{CGR, CWB} and numerous other
theoretical and experimental papers where they were discussed. If
an electron is strictly confined to a loop $\Gamma$ the effect is
contained in the dependence of the corresponding eigenvalues $E_j$
on the flux $\phi$ threading the loop (measured in the units of
flux quanta, $2\pi\hbar c|e|^{-1}$); the persistent current $I_j$
in the $j$--th state is defined as the multiple $-c\partial
E_j/\partial \phi$ of the eigenvalue derivative w.r.t. the flux.
In particular, if the particle motion on such a loop is free, we
have
 % ------------- %%
 \begin{equation} \label{ideal}
E_j(\phi) = {\hbar^2\over 2m^*} \left( 2\pi\over L\right)^2
(j+\phi)^2,
 \end{equation}
 % ------------- %%
where $L$ is the loop circumference, so the currents depend
linearly on the applied field. For a soft quantum wire in the form
of a closed curve with an attractive coupling which is constant
around the loop and strong enough the existence of persistent
currents was demonstrated in \cite{EY3}. An illustration of this
effect can be seen in a related model \cite{CE} in which a curve
supported $\delta$-coupling is replaced by an array of
two-dimensional point interaction. In analogy with this paper we
also expect that a randomization of the coupling will destroy a
coherent transport around the ring, however, since this problem is
numerically very demanding we postpone it to a subsequent paper.

%%%%%%%%%%%%%%%%%%%%%%%%%%%%%%%%%%%%%%%%%%%%%%%%%%%%%%%%%%%%%%%%%%%

\setcounter{equation}{0}
\section{Formulation of the problem}

Let $\Gamma:=\{ x\,:\; |x|=R\,\}$ with an $R>0$ be a circle in
$\mathbb{R}^2$ and $\alpha:\: \Gamma\to\mathbb{R}$ a piecewise
continuous function. We are going to discuss the Hamiltonian in
$L^2(\mathbb{R}^2)$ which can be formally written as
 % ------------- %%
 \begin{equation} \label{formalH}
H_\mathrm{form}= (-i\nabla-A(r))^2 -\alpha(\varphi) \delta(r-R)\,,
 \end{equation}
 % ------------- %%
where $r,\varphi$ are the polar coordinates of a point $x$ and $A$
is a vector potential determining the magnetic field. We suppose
that the latter is rotationally symmetric. For convenience we
associate positive values of the coupling strength $\alpha$ with
the attractive interaction. As usual we also simplify the
treatment by using rationalized units, $\hbar = 2m^*= e = c =1$.

The operator (\ref{formalH}) can be defined properly, and quite
generally, by means of the corresponding quadratic form. As long
as the function $\alpha$ is sufficiently regular, however, one use
alternatively boundary conditions \cite[Rem.~4.1]{BEKS}, i.e. to
consider the operator $H_{\alpha,A}$ acting as
 % ------------- %
 \begin{equation} \label{Ham}
 \left(H_{\alpha,A}\psi \right)(x) =
 \left((-i\nabla-A)^2\psi\right)(x)\,, \quad x\in
 \mathbb{R}^2\setminus\Gamma\,,
 \end{equation}
 % ------------- %
for any $\psi$ of the domain consisting of functions which belong
to $W_{2,2}(\mathbb{R}^2\setminus\Gamma)$, are continuous at
$\Gamma$ with the normal derivatives having a jump there,
 % ------------- %
 \begin{equation} \label{bc}
 \left.{\partial\psi\over\partial r}(x)\right|_{r=R+} \!-
 \left.{\partial\psi\over\partial r}(x)\right|_{r=R-} \!=
  -\alpha(\varphi) \psi(x)\,, \quad x=(r,\varphi)\in\Gamma\,.
 \end{equation}
 % ------------- %
It is straightforward to check that $H_{\alpha,A}$ is e.s.a., so
the Hamiltonian of our system corresponding to the formal
expression (\ref{formalH}) can be identified with its closure.

\subsection{A warm-up: constant $\delta$ interaction on a ring}

Let us consider first the simplest case when the magnetic field is
absent, the coupling strength is constant along the ring and the
interaction is attractive, $\alpha(\varphi)=\alpha>0$. We look for
the discrete spectrum which is in view of \cite[Thm.~4.2]{BEKS}
located at the negative halfline. Since the system is rotationally
symmetric we can perform the partial-wave decomposition and adopt
the following Ansatz for the eigenfunctions
 % ------------- %
 $$ %\begin{equation} \label{}
 \psi_m(r,\varphi)=  \rho_m(r)\, \e^{im\varphi}\,,\quad
 m\in\mathbb{Z}\,.
 $$ %\end{equation}
 % ------------- %
Outside the ring the solution coincides with that of  the free
Schr\"odinger equation, including the behaviour at the origin and
at infinity, i.e.
 % ------------- %
 $$ %\begin{equation} \label{}
 \rho_m(r) = \left\{ \: \begin{array}{lcl} c_1I_m(\kappa r) &
 \quad\dots\quad& r\le R \\ c_2K_m(\kappa r) &
 \quad\dots\quad& r\ge R \end{array} \right.
 $$ %\end{equation}
 % ------------- %
corresponding to the energy $-\kappa^2$. Using the continuity of
the function $\rho_m$ at $r=R$ together with the matching
condition (\ref{bc}) and the Wronskian relation $W(K_m(z),I_m(z))=
z^{-1}$ we get the spectral condition
 % ------------- %
 \begin{equation} \label{fullring}
 (I_mK_m)(\kappa R)= {1\over\alpha R}\,.
 \end{equation}
 % ------------- %
To find its solutions, recall some properties of the modified
Bessel functions \cite[Chap.~9]{AS}; since $I_{-m}(z)=I_m(z)$ and
$K_{-m}(z)=K_m(z)$, we may consider $m\ge 0$ only. The l.h.s. of
(\ref{fullring}) is decreasing in $(0,\infty)$ and has the
following asymptotics,
 % ------------- %
 $$ %\begin{equation} \label{}
 (I_mK_m)(z)= \left\{ \: \begin{array}{lcl} -\ln{z\over 2}
 -\gamma + o(1) &
 \quad\dots\quad& m=0 \\ {1\over 2m}+ \mathcal{O}(z) &
 \quad\dots\quad& m\ne 0 \end{array} \right.
 $$ %\end{equation}
as $z\to 0$, where $\gamma=0.577\dots$ is the Euler number, and
 % ------------- %
 $$ %\begin{equation} \label{}
 (I_mK_m)(z)= {1\over 2z}\, \left\lbrack\, 1- {2\left(m^2\!
 -{1\over 4}\right)\over (2z)^2} + \mathcal{O}(z^{-3})
 \,\right\rbrack
 $$ %\end{equation}
 % ------------- %
as $z\to\infty$. Consequently, there is at most one solution at each
partial wave which we have anticipated labeling the solution by
the orbital index only. The ground state corresponding to $m=0$
exists for any $\alpha>0$, while other partial waves exhibit a
bound state provided
 % ------------- %
 $$ %\begin{equation} \label{}
 \alpha R> 2|m|\,.
 $$ %\end{equation}
 % ------------- %
In the weak-coupling case, $\alpha R\ll 1$, the binding energy is
exponentially small,
 % ------------- %
 $$ %\begin{equation} \label{}
 E_0\approx -{4\over R^2}\, \e^{-2/\alpha R},
 $$ %\end{equation}
 % ------------- %
as expected for a two-dimensional Schr\"odinger
operator\footnote{Taking the limit $R\to 0$ with a properly scaled
$\alpha(R)$ we can get a two-dimensional point interaction. If
$\alpha$ depends on the angle, such a limit can be done using the
resolvent formula of \cite{BEKS} in analogy with the
tree-dimensional case worked out recently in \cite{Sh}.}, while
for $\alpha R\gg 1$ the above asymptotics gives
 % ------------- %
 $$ %\begin{equation} \label{}
 E_m= -{\alpha^2\over 4} + {{m^2-{1\over 4}}\over R^2}
 + \mathcal{O}(\alpha^{-2}R^{-4})\,.
 $$ %\end{equation}
 % ------------- %
This can be regarded as a particular case of the theorem proven in
\cite{EY1}, just the error estimate is better than in the general
case. The discrete spectrum for a fixed coupling strength as a
function of $R$ is plotted in Fig.~1.

One can continue the analysis of the rotationally symmetric
$\delta$-interaction on the ring by discussing the scattering,
approximations by means of scaled potentials, combinations with
other interactions, etc., in analogy in the three-dimensional case
considered in \cite{AGS}. We will not do that, however, because we
are interested mainly in the situation without the rotational
symmetry.

\subsection{General formalism}

Next we want to look what happens if the rotational symmetry is
broken due to non-constancy of the function $\alpha(\cdot)$. In
this case the point interaction couples different partial waves
and we have to look for the solution in the form of a series.
Since the scheme is similar for the magnetic case, or even for
operators
 % ------------- %
 $$ %\begin{equation} \label{}
 H_{\alpha,A} + V(r)
 $$ %\end{equation}
 % ------------- %
with a rotationally invariant potential, we shall formulate it
generally adopting the following Ansatz
 % ------------- %
 \begin{equation} \label{Ansatz}
 \psi(r,\varphi) = \left\{ \: \begin{array}{lcl}
 \sum_{m\in\mathbb{Z}}c_m f_m(r)\, \e^{im\varphi} &
 \quad\dots\quad& r\le R \\ \sum_{m\in\mathbb{Z}}d_m
 g_m(r)\, \e^{im\varphi} &
 \quad\dots\quad& r\ge R \end{array} \right.
 \end{equation}
 % ------------- %
where $\{c_m\},\: \{d_m\}$ are coefficient sequences and
$f_m,\,g_m$ are solutions of the ``free'' Schr\"odinger equation
with the energy $-\kappa^2$ inside and outside the ring,
respectively, in the $m$-th partial wave; recall that by
assumption the system without the point interaction is
rotationally symmetric. We impose the standard requirement that
the solution is regular at the origin and $L^2$ at infinity, then
the functions $f_m,\,g_m$ are unique up to multiplicative
constants which can be absorbed into the coefficients.

The above function has naturally to belong to the domain of the
Hamiltonian. Its continuity at $r=R$ together with the
orthonormality of the trigonometric basis in $L^2((0,2\pi))$
implies
 % ------------- %
 \begin{equation} \label{cont}
 c_m = d_m\, {g_m(R)\over f_m(R)}\,.
 \end{equation}
 % ------------- %
At the same time, the matching condition (\ref{bc}) yields
 % ------------- %
 $$ %\begin{equation} \label{}
 \sum_m d_m
 g'_m(R)\, \e^{im\varphi}
 - \sum_m c_m
 f'_m(R)\, \e^{im\varphi}
 = -\alpha(\varphi)\sum_m d_m
 g_m(R)\, \e^{im\varphi}\,;
 $$ %\end{equation}
 % ------------- %
multiplying this relation by $\e^{-in\varphi}$ and integrating
w.r.t. $\varphi$ we get an expression of $c_n$ by means of the
coefficients $d_m$. Combing it with (\ref{cont}) and introducing
 % ------------- %
 $$ %\begin{equation} \label{}
 \alpha_{mn} := \int_0^{2\pi} \alpha(\varphi)\,
 \e^{i(m-n)\varphi}\,\D\varphi\,, \quad
 \langle\alpha\rangle := \alpha_{mm} = \int_0^{2\pi}
 \alpha(\varphi)\,\D\varphi
 $$ %\end{equation}
 % ------------- %
we get after a simple manipulation
 % ------------- %
 \begin{equation} \label{derijump}
 \sum_{m\in\mathbb{Z}} \Big\lbrace \delta_{mn} \left\lbrack
 \langle\alpha\rangle f_n g_n + 2\pi W(f_n,g_n) \right\rbrack(R) +
 (1\!-\!\delta_{mn}) \alpha_{mn} (g_m f_n)(R) \Big\rbrace
 d_m = 0
 \end{equation}
 % ------------- %
as a set of equations for the coefficients, or in other words, an
operator equation in the $\ell^2$ space of the coefficients.
Solving it numerically, one takes a family of truncated systems;
the convergence of such an approximation is checked in the same
way as in \cite{ESTV}.

\subsection{A broken ring}

Consider now the case when the coupling strength is a step
function,
 % ------------- %
 $$ %\begin{equation} \label{}
 \alpha(\varphi) = \alpha \chi_{[\theta/2,2\pi-\theta/2]}
 (\varphi)
 $$ %\end{equation}
 % ------------- %
for some $\alpha>0$ and $\theta\in (0,2\pi)$. Substituting
$W(I_m(\kappa R),K_m(\kappa R))= -R^{-1}$ (recall that the
derivative is taken w.r.t. $r$) and
 % ------------- %
 $$ %\begin{equation} \label{}
 \alpha_{mn} = \left\{ \: \begin{array}{lcl}
 \alpha(2\pi-\theta) & \quad\dots\quad& m=n \\
 -{2\alpha\over m\!-\!n}\, \sin {\theta(m\!-\!n)
 \over 2} & \quad\dots\quad& m\ne n \end{array} \right.
 $$ %\end{equation}
 % ------------- %
into (\ref{derijump}) we get equations for the coefficients $d_m$
in this case.

In an ideal quantum wire described by an appropriate Schr\"odinger
operator on $\Gamma$ it is the topology rather than geometry which
determines the character of the discrete spectrum. For the full
ring, $\theta=0$, the spectrum is twice degenerate with the
exception of the ground state, while a broken ring spectrum looks
like that of an infinite deep rectangular well. The spectrum of a
soft wire will be similar if the distance between the loose ends
is large so that the tunneling between them is negligible. To get
an idea at what values of the parameters the regime changes,
recall that the eigenfunction of the one-center point interaction
of the coupling strength $-\alpha$ on line is $\e^{-\alpha
|x|/2}\:$ \cite{AGHH}, and thus the characteristic ``size'' of the
bound-state wave function measured from the ring is
$2\alpha^{-1}$. Comparing this to the ``gap size'' we see that the
switch from the full-ring-type spectrum to the infinite-well-type
takes place at $\alpha\approx 2\theta R$. This is illustrated in
Fig.~2 where we plot the spectrum as a function of the gap angle;
it is seen that the transition takes place around the value where
we expect it.

%%%%%%%%%%%%%%%%%%%%%%%%%%%%%%%%%%%%%%%%%%%%%%%%%%%%%%%%%%%%%%%%%%%

\setcounter{equation}{0}
\section{Magnetic rings}

\subsection{The homogeneous background field}

Consider the magnetic field of intensity $B>0$ perpendicular to
the ring plane. In the circular gauge the corresponding vector
potential is $A= {1\over 2}\, Br\, e_{\varphi}$ and the
Hamiltonian without the singular interaction at the ring acts as
 % ------------- %
 $$ %\begin{equation} \label{}
 H_{0,A} = -{\partial^2\over\partial r^2} - {1\over r}
 {\partial\over\partial r} + {1\over r^2} \left(
 -i{\partial\over\partial\varphi} + {1\over 2}\, Br^2 \right)^2 .
 $$ %\end{equation}
 % ------------- %
After a standard manipulation we find that the solutions at energy
$z$ for an interior and exterior of the ring, to be inserted into
(\ref{Ansatz}), are
 % ------------- %
 \begin{eqnarray*} \label{}
 f_m(r) &\!=\!& r^{|m|}\, \e^{-Br^2/4}\, M\left( a_m(z),
 |m|+1; {1\over 2}\, Br^2 \right)\,, \\
 g_m(r) &\!=\!& r^{|m|}\, \e^{-Br^2/4}\, U\left( a_m(z),
 |m|+1; {1\over 2}\, Br^2 \right)\,,
 \end{eqnarray*}
 % ------------- %
respectively, where
 % ------------- %
 $$ %\begin{equation} \label{}
 a_m(z):= {1\over 2}\, \left( m+|m|+1 - {z\over B} \right)
 $$ %\end{equation}
 % ------------- %
and $M,\, U$ are the regular and singular confluent hypergeometric
function, respectively. In particular, the spectrum of the free
(Landau) Hamiltonian is given by the condition $a_m(z)= -n,\:
n=0,1,2,\dots$, i.e. consists of the Landau levels
 % ------------- %
 $$ %\begin{equation} \label{}
 z = B(m+|m|+2n+1)\,, \quad n=0,1,2,\dots\,.
 $$ %\end{equation}
 % ------------- %
To employ (\ref{derijump}) we need also the Wronskian of the above
solutions which is by \cite[13.1.22]{AS} equal to
 % ------------- %
 $$ %\begin{equation} \label{}
 W(f_m(r),g_m(r)) = -\,{2\over r} {\Gamma(|m|+1)\over \Gamma(a_m(z))}\, \left(
 {B\over 2 }\,  \right)^{-|m|} .
 $$ %\end{equation}
 % ------------- %
In particular, if the ring is rotationally symmetric,
$\alpha(\varphi)=\alpha$, the condition (\ref{derijump}) is
reduced to
 % ------------- %
 $$ %\begin{equation} \label{}
 {\left( {1\over 2}\, BR^2 \right)^{|m|} \over 2|m|!}\,
 \Gamma(a_m(z)) \e^{-BR^2/2} (MU)\left( a_m(z), |m|+1; {1\over 2}\, BR^2
 \right) = {1\over\alpha R}\,.
 $$ %\end{equation}
 % ------------- %

\subsection{The AB flux case}

Suppose now that instead of the homogeneous field considered above
the ring is threaded at its centre by a magnetic flux line. The
vector potential can be now chosen as $A= {\Phi\over 2\pi r}\,
e_{\varphi}$, where $\Phi$ is the value of the flux. The
normalized flux appearing in (\ref{ideal}) equals in the present
units $\phi= \Phi/2\pi$.

We assume that the flux line is not combined with a point
interaction \cite{AT, DS}; treating then the Schr\"odinger
equation in a standard way \cite{Ru} we get the interior and
exterior solutions at the energy $-\kappa^2$ in the form
 % ------------- %
 $$ %\begin{equation} \label{}
 f_m(r) = I_{|m-\phi|}(\kappa r)\,, \quad
 g_m(r) = K_{|m-\phi|}(\kappa r)\,,
 $$ %\end{equation}
 % ------------- %
respectively; their Wronskian is
 % ------------- %
 $$ %\begin{equation} \label{}
 W(f_m(r),g_m(r)) = -\, {1\over r}\,.
 $$ %\end{equation}
 % ------------- %
In particular, in case of a rotationally symmetric ring we get the
spectral condition which is similar to (\ref{fullring}), namely
 % ------------- %
 \begin{equation} \label{fullmgring}
 (I_{|m-\phi|} K_{|m-\phi|})(\kappa R)= {1\over\alpha R}\,.
 \end{equation}
 % ------------- %
Taking into account the relation
 % ------------- %
 $$ %\begin{equation} \label{}
 \lim_{z \rightarrow 0} (I_{|\nu|} K_{|\nu|})(z) = {1\over 2|\nu|}
 $$ %\end{equation}
 % ------------- %
we can find the condition
 % ------------- %
 \begin{equation} \label{ABex}
 2|m-\phi| = \alpha R
 \end{equation}
 % ------------- %
that determines critical values at which the eigenvalues are
absorbed in the essential spectrum.

%%%%%%%%%%%%%%%%%%%%%%%%%%%%%%%%%%%%%%%%%%%%%%%%%%%%%%%%%%%%%%%%%%%

\setcounter{equation}{0}
\section{Results and discussion}

Let us now turn to a more detailed discussion of the results.

\subsection{Non-magnetic case, constant $\alpha$}

We have presented already in Figs.~1,2 the spectrum for the full
and broken ring, respectively. If the radius is large enough (in
terms of the quantity $\alpha R$) the spectrum has a distinctive
one-dimensional character with the largest density at its bottom.
At the same time, a gap which is large enough to prevent tunneling
between the loose ends gives the spectrum the hard-wall nature.
This leads us to the \emph{conjecture} that the asymptotic formula
derived in \cite{EY1} will hold for any finite, smooth, and
\emph{non-closed} curve as well, with the comparison operator $S=
-{d^2\over ds^2} - {1\over 4}k(s)^2$, where $k(s)$ is the
curvature, being now specified by \emph{Dirichlet} boundary
conditions at the curve endpoints. The one-dimensional character
can be seen also from the eigenfunctions: in Fig.~3 we plot the
ground state and the eighth excited state.

\subsection{Non-magnetic case, localization}

As we have said in the introduction one expects that irregular
variations of the coupling constant may destroy the coherence of
the wavefunction along the curve supporting the interaction. Since
the negative spectrum of the present model is discrete in any
case, one can try to see the effect through the characteristic
size of the eigenfunctions. If the coupling strength $\alpha$ is
constant over the whole ring or a substantial part of it, the
latter are distributed roughly uniformly along $\Gamma$; the
uniformity is ideal for the full ring and it has sinusoidal
variations due to the natural quantization for excited states.

When we want to find a quantity to characterize the ``size'' of
the eigenfunction $\psi_k$ corresponding to the energy $E_k$ we
notice first that it can be expressed as \cite{Po}
 % ------------- %
 $$ %\begin{equation} \label{}
 \psi_k(x) = \int_\Gamma \psi(y(\varphi)) G_0(x,y(\varphi); E_k)\,
 \D\varphi\,,
 $$
 % ------------- %
possibly up to a normalization factor, where $y(\varphi)$ is a
point of the ring and $G_0$ is the Green function of the
two-dimensional Laplacian. Consequently, it is sufficient to
characterize $\psi_k$ by its restriction to $\psi_k(\varphi)
\equiv \psi_k(y(\varphi))$ to the ring. Then we choose as the
quantity of interest the second moment of the corresponding
probability distribution, minimized over the choice of the
reference point,
 % ------------- %
 $$ %\begin{equation} \label{}
 \langle\Delta\psi\rangle := \min_{\varphi_0} \left( \int_0^{2\pi}
 (\varphi-\varphi_0)^2
 |\psi_k(\varphi)|^2 \D\varphi \right)^{1/2}.
 $$
 % ------------- %
Now we take a family of random step functions $\alpha(\cdot)$ and
plot the above quantity for the ground-state eigenfunction in
dependence on the dispersion $\Delta\alpha$; the result is plotted
in Fig.~4. It is obvious that in the average a stronger coupling
randomness means a ``more localized'' eigenfunction.

\subsection{Magnetic rings}

Let us pass now to the case with a non-zero magnetic field. If the
latter in homogeneous, the spectrum is pure point and it
accumulates at the Landau levels. In Fig.~5 we show the
eigenvalues belonging to the three lowest ``bands'' of the full
ring as functions of the magnetic field intensity. Fig.~6 shows
the situation in a cut ring with $B$ fixed and the gap angle
$\theta$ changing. At the limit $\theta\to 2\pi$ the eigenvalues
are, of course, absorbed in the Landau levels. What is more
interesting are the avoided crossings in the higher ``bands''
which witness about the non-trivial character of the cyclotronic
motion around such an ``obstacle''. Another illustration of the
fact is given by the eigenfunction contour plots shown in Fig.~7.

On the other hand, in the Aharonov-Bohm case the essential
spectrum covers the positive halfline, and the dependence of the
negative eigenvalues on the magnetic flux shown in Fig.~8
resembles a picture corresponding to (\ref{ideal}) for the ideal
ring graph, that is a family of shifted parabolas. The same can be
said, of course, about the lower part of the homogeneous-field
spectrum in Fig.~5, but in distinction to that case the AB
spectrum is periodic modulo an integer number of flux quanta. A
common feature is a strong dependence of the eigenvalues on the
flux, which by the formula given in the introduction means that
such system exhibit non-negligible persistent currents.

Recall in this connection a close relation between such
current-carrying states and the edge currents \cite{Ha, MS}.
Furthermore, to create a magnetic transport one does not need a
hard wall; much more gentle perturbations like a potential
``ditch'', translationally invariant field modification \cite{Iw,
MP, EK} or even an array of point obstacles \cite{EJK1} are
sufficient. If a ``linear'' perturbation is replaced by a circular
one, one can expect occurrence of localized states carrying
current along the circle. Naturally, the magnetic transport may be
destroyed by a disorder. The problem of stability of (a part of)
absolutely continuous spectrum with respect to perturbations has
been studied recently for edge currents in halfplanes and similar
domains \cite{BP, FGW, MMP}, strips \cite{EJK2} and compact
domains \cite{FM}. In a finite-length setting the localization due
to disorder was shown in \cite{CE}; we intend to do the same in
the context of the present model in a subsequent publication.

While two types of the magnetic field yield similar results, there
are important differences. One of them concerns the character of
the perturbation. The Aharonov-Bohm Hamiltonian is \emph{not} an
analytic perturbation of the operator with $\phi=0$. This is
obvious from the fact that the eigenvalue curves are in this case
non-smooth at the integer values of the flux $\phi$ as one can see
in Figs.~9 and 10 showing the spectrum for different $\alpha$'s
and gap angles\footnote{The same is true in Fig.~8 where, however,
the jump in the derivative is too small to be visible.}; in the
full-ring case we present a comparison with the homogeneous field
having the same flux through $\Gamma$. Another difference from the
homogeneous field case is that the discrete spectrum may be void
for some values of the parameters which is again clear from the
said figures; for the full ring the critical value is given by
(\ref{ABex}).

\subsection*{Acknowledments}

The research has been partially supported by GAAS grant A1048101.

\newpage

\subsection*{Figure captions}

\begin{description}
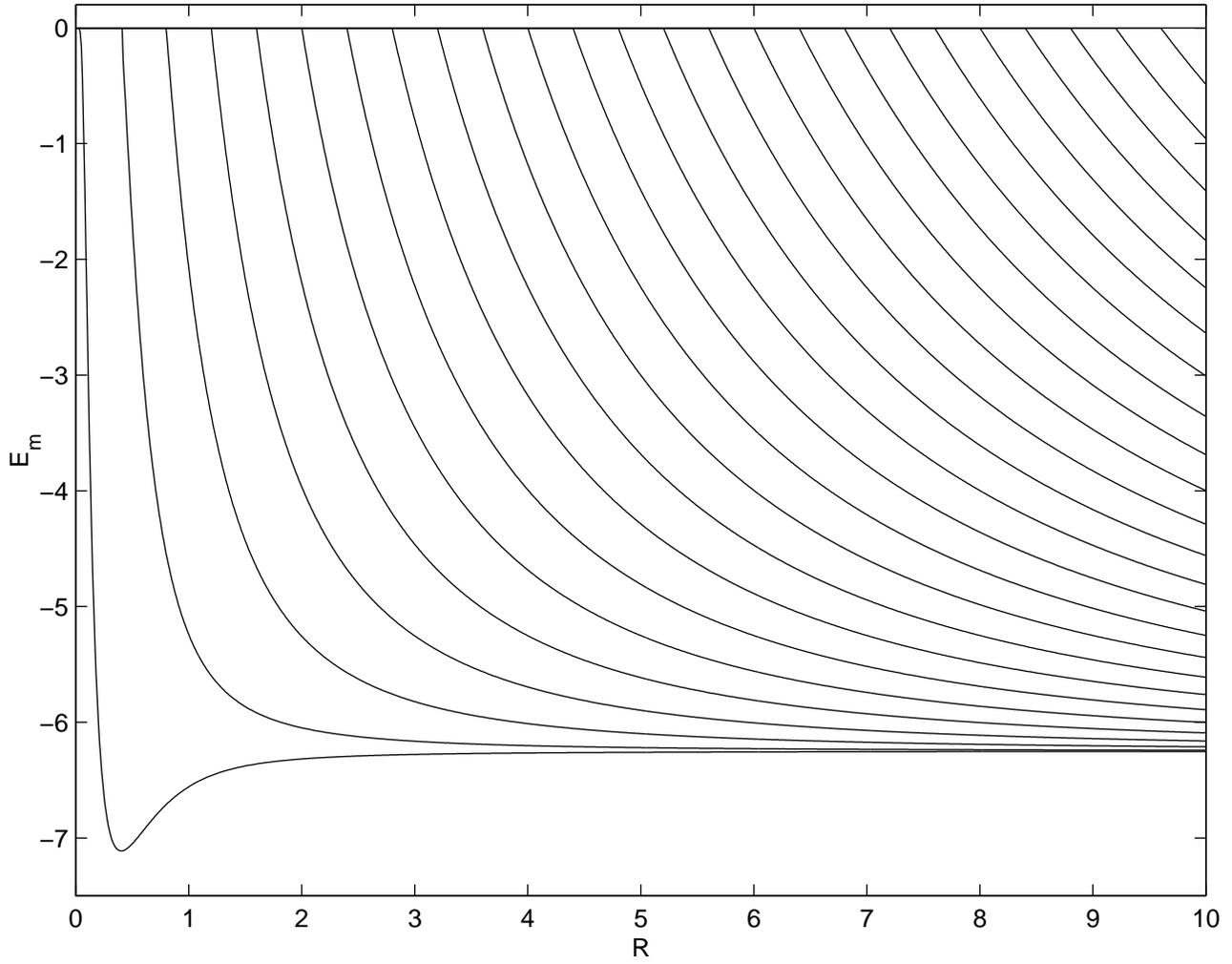
   \item{\bf Figure 1\quad} The dependence of energy eigenvalues of a full
ring on the radius $R$, $\alpha=5$. Each level is labelled by $m$,
all levels except the ground state are twice degenerated.
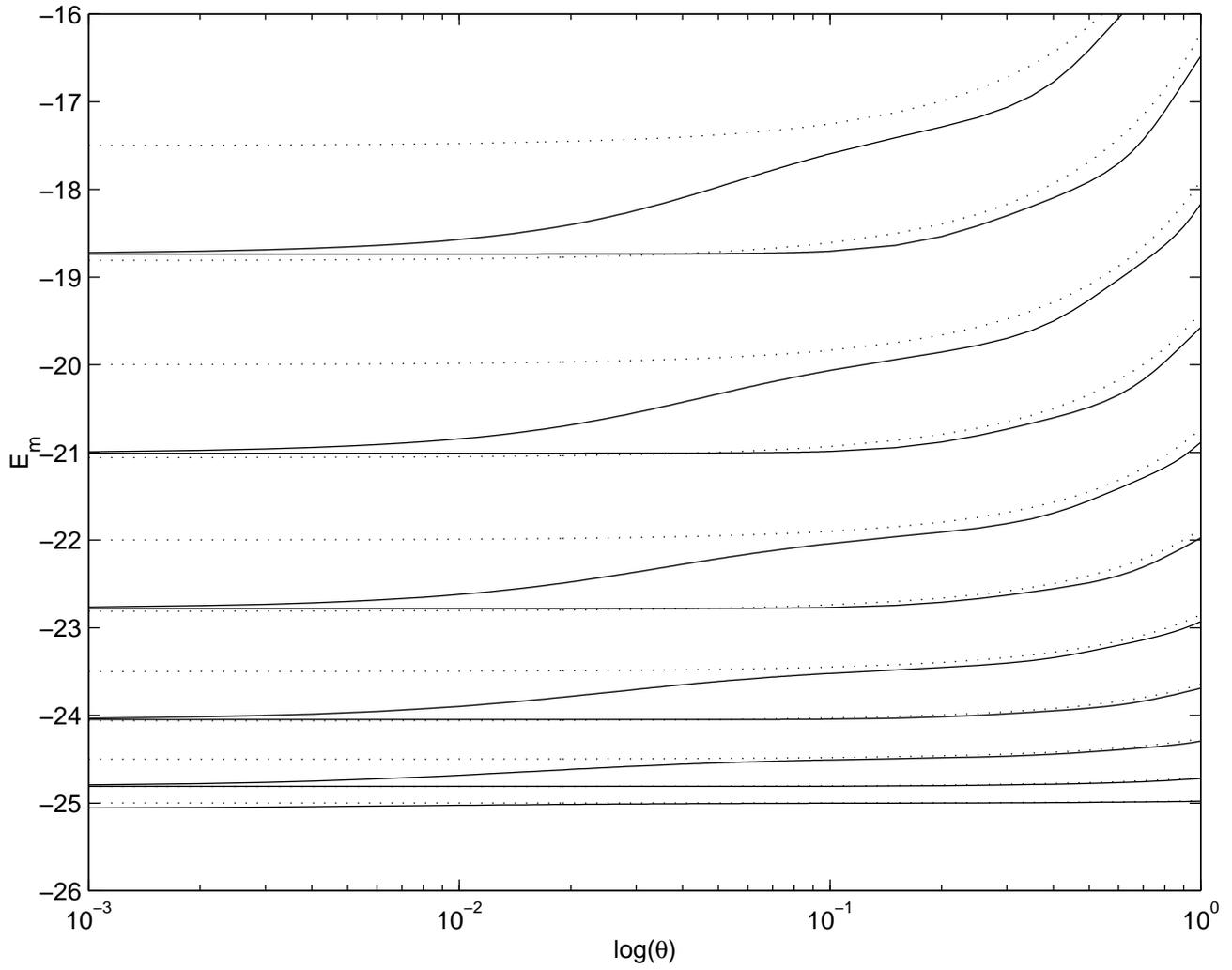
   \item{\bf Figure 2\quad} This figure shows a transition between the
two regimes as the distance $\theta$ between the loose ends of the
ring increases. The full-ring-type spectrum $(\theta\approx0)$
passes to the infinite-well-type around $\alpha\approx 2\theta R$.
In the presented situation $R=2$ and $\alpha=10$ is chosen. The
dotted lines show the spectrum of a Dirichlet case.
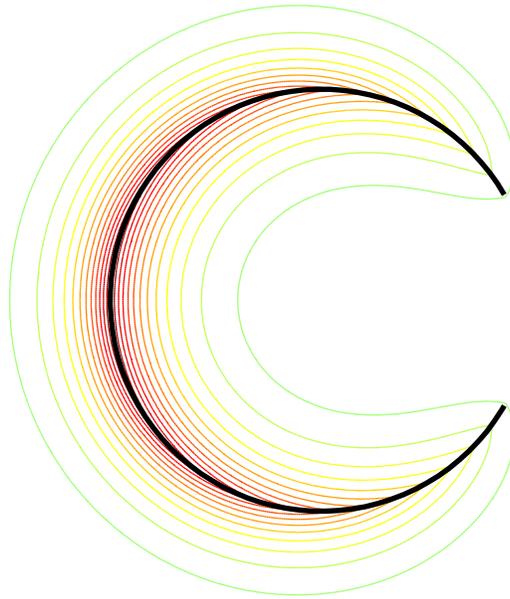
   \item{\bf Figure 3\quad} The contour plots of the (real) wave functions
of the ground state and the $8th$ excited states of a broken ring
are shown for $\alpha=1$, $R=10$, and $\theta=\pi/3$. The
corresponding eigenenergies are $E_0=-0.249$ and $E_8=-0.00415$,
respectively.
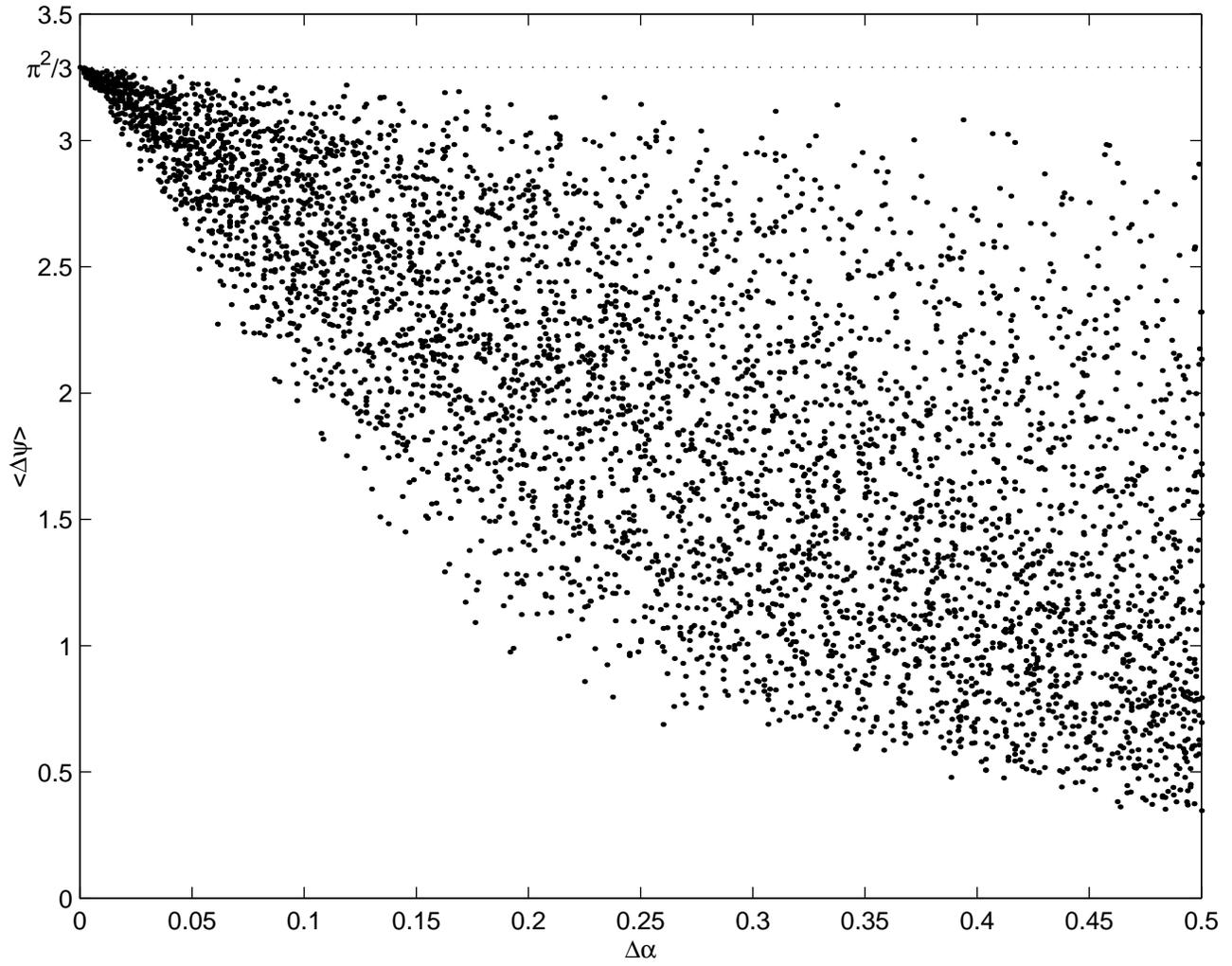
   \item{\bf Figure 4\quad} The influence of random choice of
$\alpha$ on localization of the ground-state eigenfunction is
presented. Each of the 5000 points represents a particular
(random) choice of the piecewise constant function
$\alpha(\varphi)$ that is formed by 10 constant pieces of the same
angle length $\pi/5$ and the value of $\alpha$ is uniformely
distributed around $\alpha_0=1$ with a given mean square
$\Delta\alpha$. The quantity $\langle\Delta\psi\rangle$ is the
(minimized) second moment of the probability distribution. The
radius $R=5$.
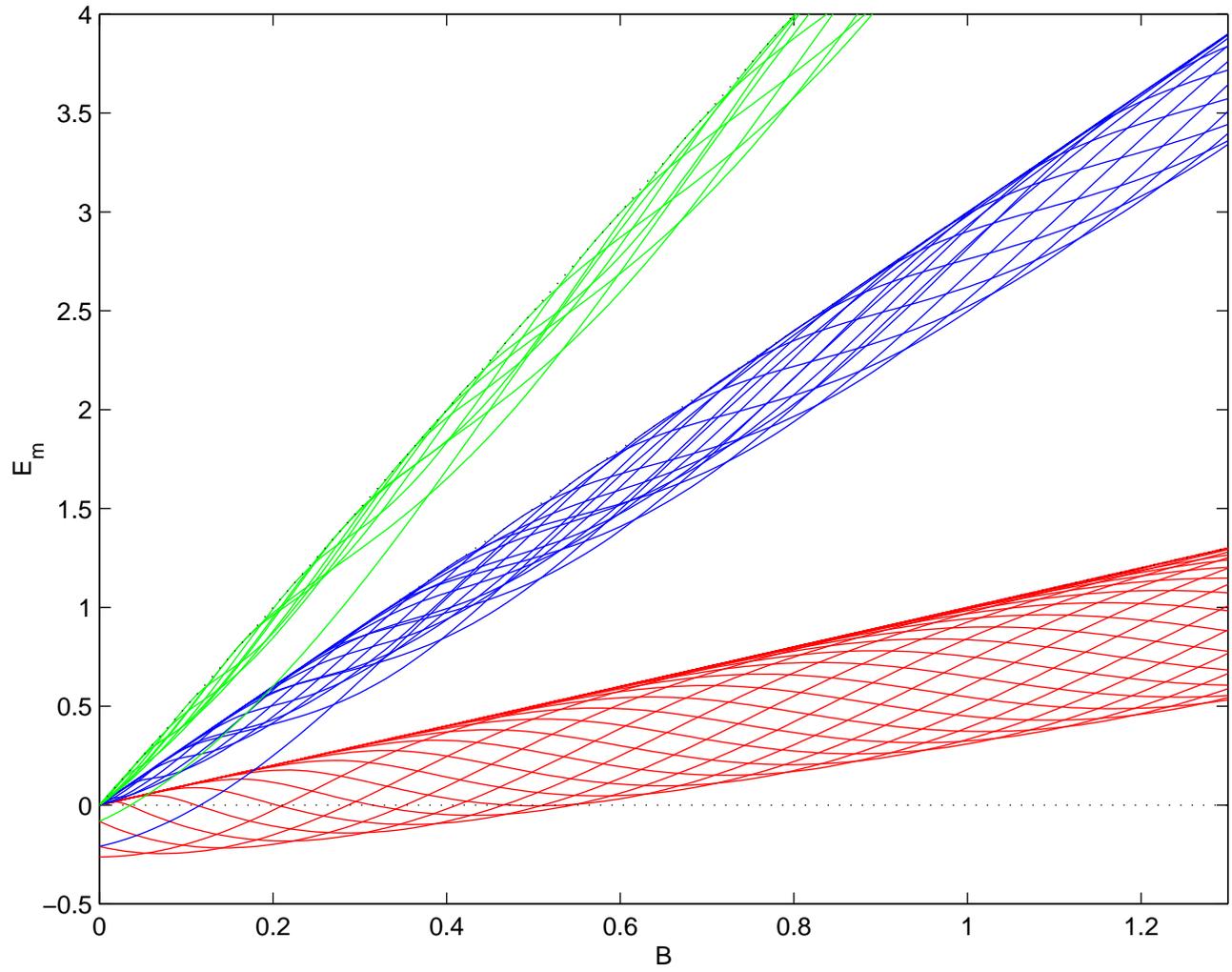
   \item{\bf Figure 5\quad} The dependence of energy eigenvalues of a full ring
on intensity of homogeneous magnetic field $B$, $\alpha=1$, $R=5$.
We plot only levels ``setting on" the first three Landau levels.
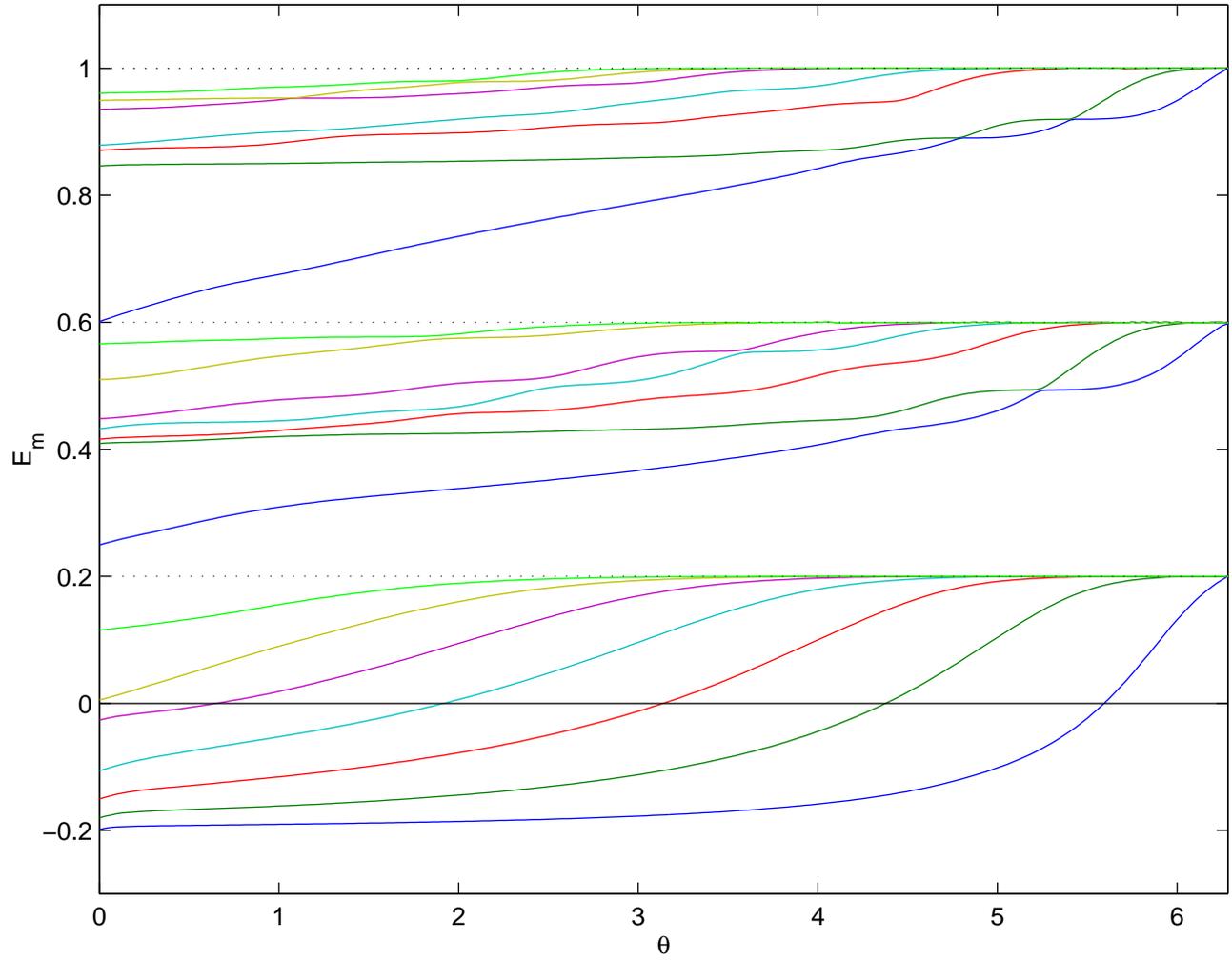
   \item{\bf Figure 6\quad} The dependence of energy eigenvalues of a broken
ring on the gap angle, $B=0.2$, $\alpha=1$, $R=5$. Only first
seven levels of each ``band" are plotted. The dotted lines mark
the Landau levels for the given $B$.
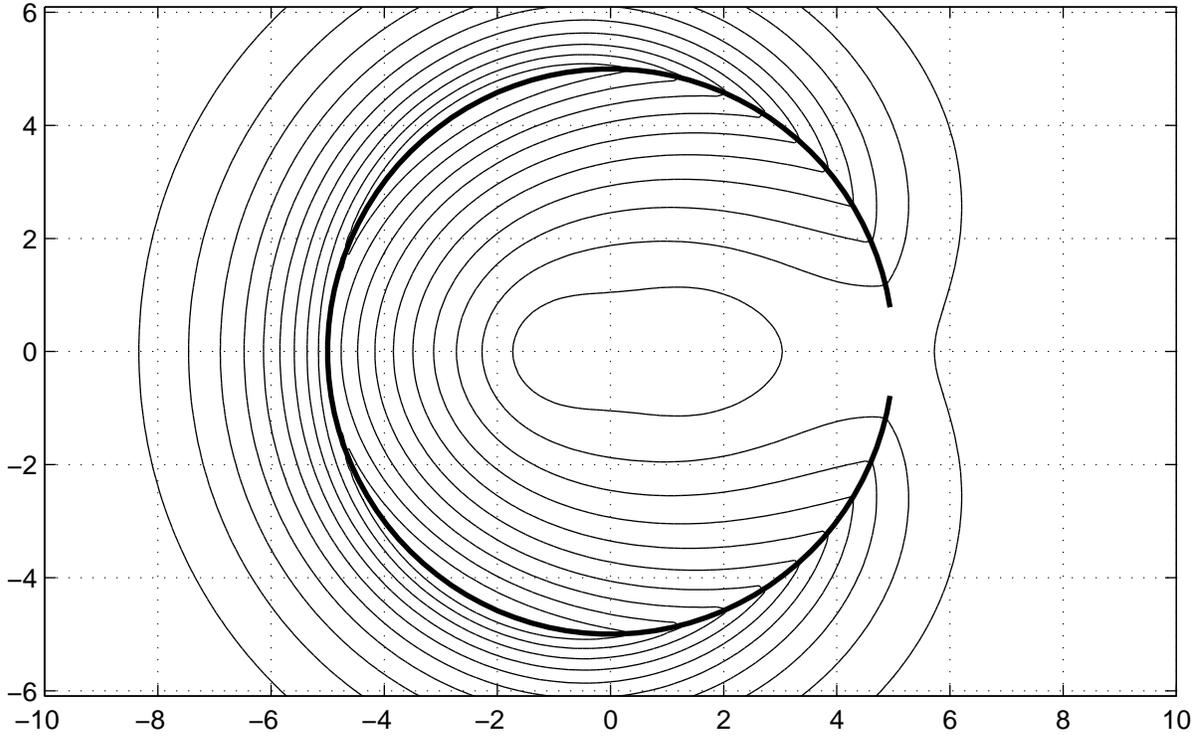
   \item{\bf Figure 7\quad} The contour plots of the absolute value of wave
functions of broken rings, $B=0.2$, $\alpha=1$, $R=5$. The upper
subplot shows the ground state $(E=-0.193)$ of a ``almost full"
ring $\theta=\pi/10$ and the lower one an excited state
$(E=0.552)$ of the second ``band" with $\theta=\pi$.
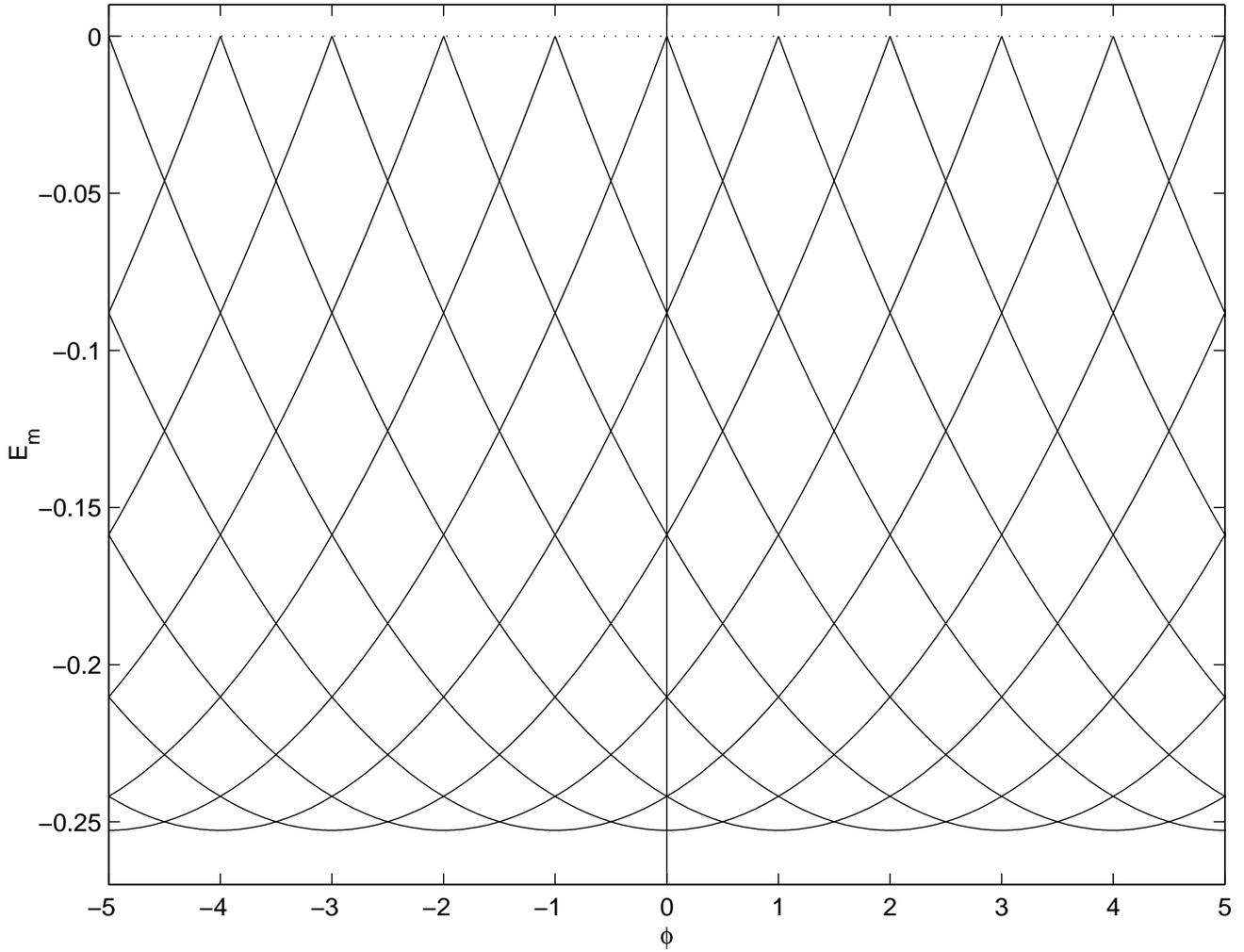
   \item{\bf Figure 8\quad} The dependence of energy eigenvalues of a full
ring in the Aharonov-Bohm case on the magnetic flux $\phi$,
$\alpha=1$, $R=10$.
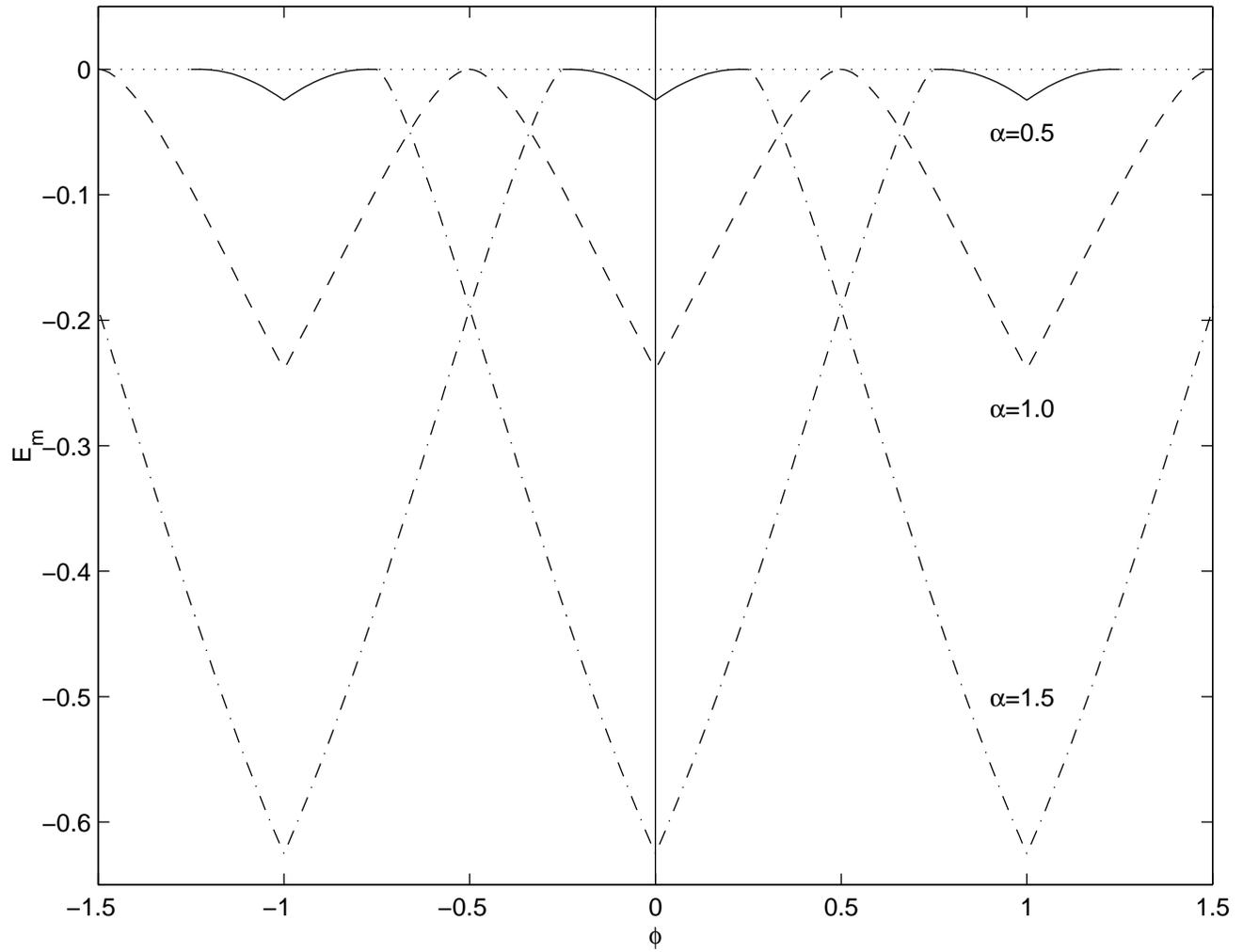
   \item{\bf Figure 9\quad} The dependence of energy eigenvalues of a full
ring in the Aharonov-Bohm case on the magnetic flux $\phi$, $R=1$.
The three values of $\alpha=0.5, 1, 1.5$ are chosen so as to
demonstrate both the non-smootheness of energy curves at integer
values of $\phi$ and the possibility of a void spectrum for some
intervals of $\phi$.
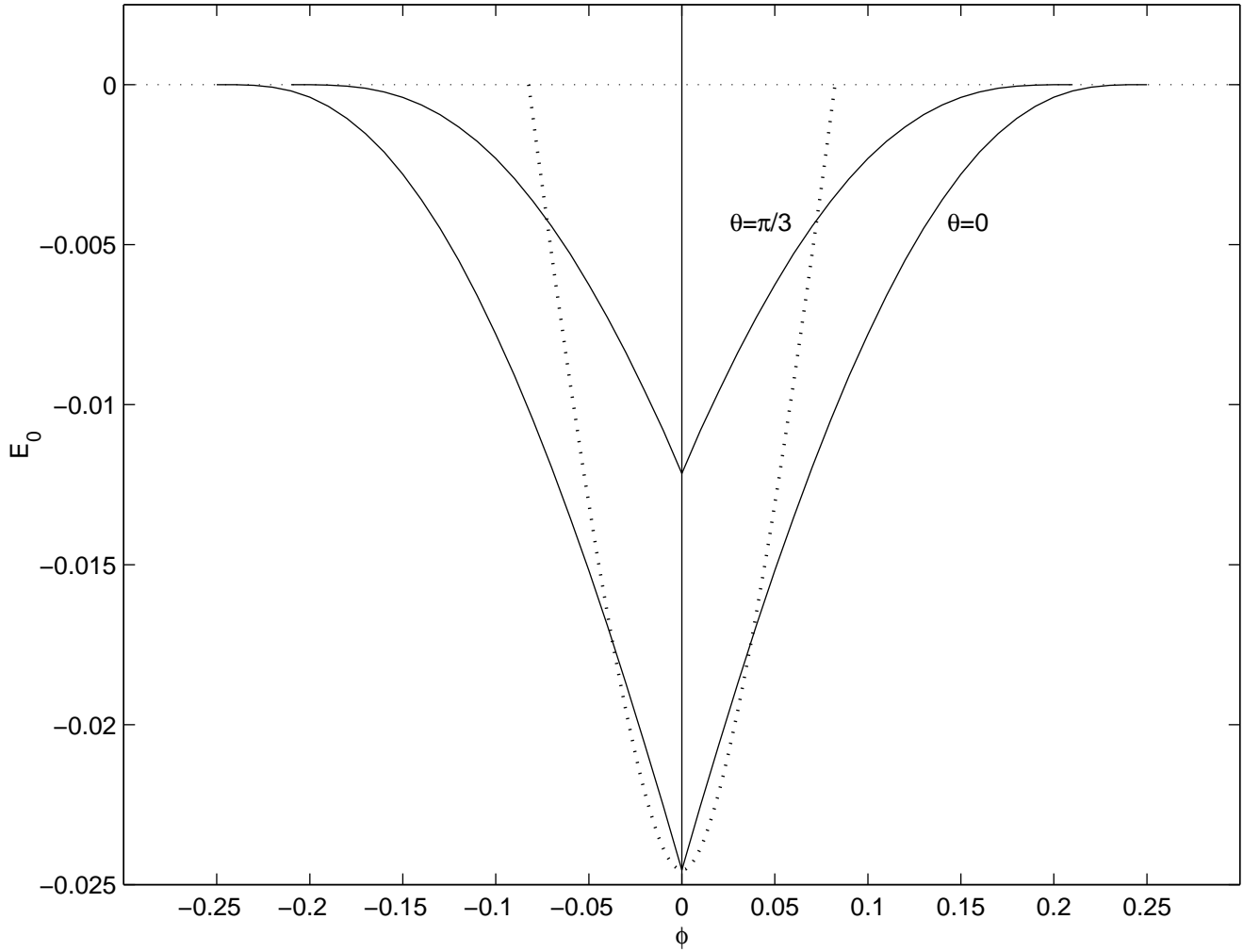
   \item{\bf Figure 10\quad} The comparison of dependence of eigenenergies
on the magnetic flux $\phi$ for a full $\theta=0$ and broken
$\theta=\pi/3$ rings in the Aharonov-Bohm case, $\alpha=0.5$,
$R=1$. The dotted curve shows an analogical situation with the
homogeneous magnetic field of the same flux.
   \end{description}

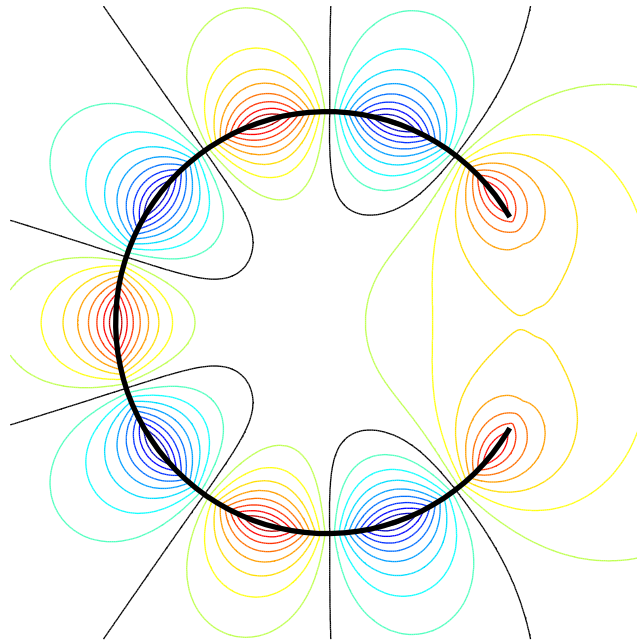
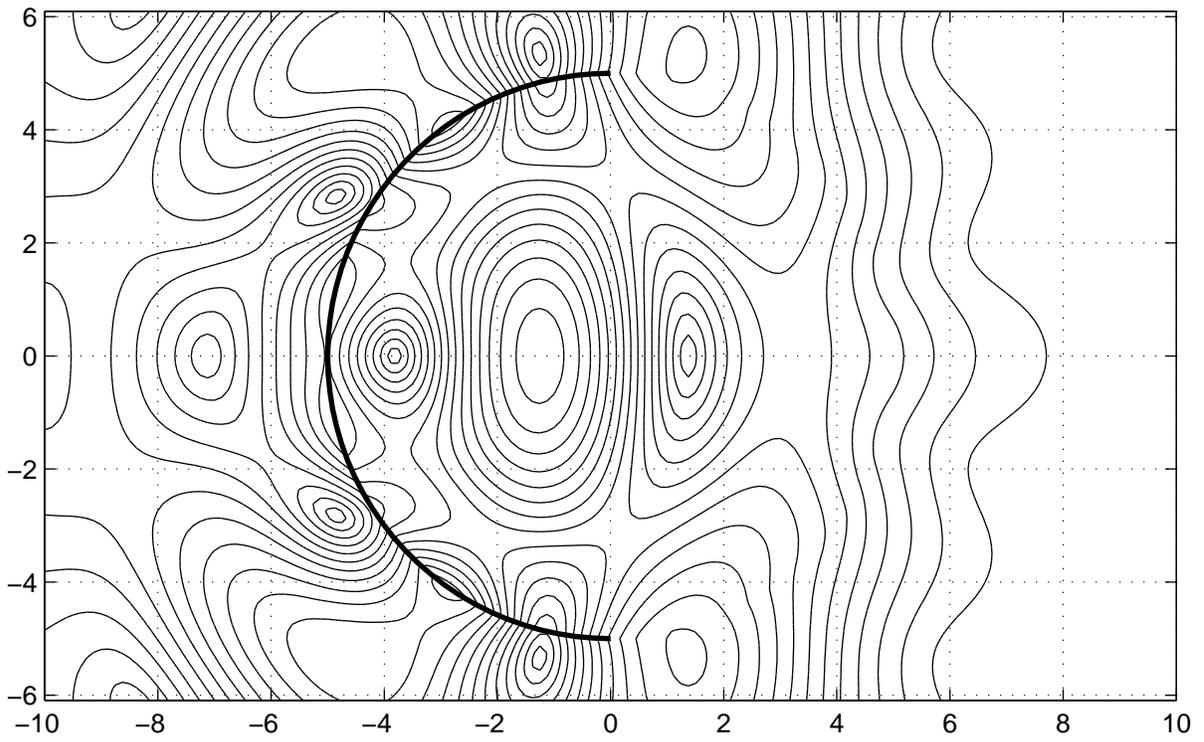
\end{document}